\documentclass[onecolumn,showpacs,preprintnumbers,amsmath,amssymb]{revtex4}

\usepackage{graphicx}
\usepackage{epsfig}

\begin{document}
\title{\bf The Dirichlet Casimir Energy for $\phi^4$ Theory in a Rectangle}
\author{M. A. Valuyan}
\email{m-valuyan@sbu.ac.ir; m.valuyan@semnaniau.ac.ir}
\affiliation{Department of Physics, Semnan Branch, Islamic Azad University, Semnan, Iran}
\date{\today}

\begin{abstract}
In this article, we present the zero and first-order radiative correction to the Dirichlet Casimir energy for massive and massless scalar field confined in a rectangle. This calculation procedure was conducted in two spatial dimensions and for the case of the first-order correction term is new. The renormalization program that we have used in this work, allows all influences from the dominant boundary conditions\,(\emph{e.g.} the Dirichlet boundary condition) be automatically reflected in the counterterms. This permission usually makes the counterterms position-dependent. Along with the renormalization program, a supplementary regularization technique was performed in this work. In this regularization technique, that we have named Box Subtraction Scheme\,(BSS), two similar configurations were introduced and the zero point energies of these two configurations were subtracted from each other using appropriate limits. This regularization procedure makes the usage of any analytic continuation techniques unnecessary. In the present work, first, we briefly present calculation of the leading order Casimir energy for the massive scalar field in a rectangle via BSS. Next, the first order correction to the Casimir energy is calculated by applying the mentioned renormalization and regularization procedures. Finally, all the necessary limits of obtained answers for both massive and massless cases are discussed.
\end{abstract}

\maketitle

\section{Introduction}
\label{sec:intro}
The concept of the Casimir effect was firstly disseminated by discussing an attractive force between two plates when placed close juxtaposed. The relevance of the Casimir effect has increased over the decades since the work of H.B.G Casimir\,(1948), who was the first to predict and explain this effect as a change in vacuum quantum fluctuations of the electromagnetic field\,\cite{h.b.g.}. This prediction was firstly examined in 1958 by M. J. Sparnaay\,\cite{Sparnaay.M.J.}. Later on, this effect received increasing attention and applications in many fields of physics\,\cite{quantum.field.theory.1,condensed.matter.1,atomic.molecular.1,astro.physics.1,Mathematical.physics.1}. After the first attempt to the calculation of the Casimir energy in the interacting quantum field theory by Bordag et al.\,\cite{Bordag.et.al}, extensive investigations were conducted on the radiative correction to the Casimir energy for multiple fields and geometries\,\cite{RC.free.counterterms.}. In the smaller category of works, for the case of real massive scalar field\,(in $\phi^4$  self-interaction theory), the two-loop correction for the Casimir energy was computed in\,\cite{13-20reza,cavalcanti.}. Moreover, N. Graham et al. introduced new approaches to this problem by utilizing the phase shift of the scattering states\,\cite{21-reza}, or replacing the boundary conditions by an appropriate potential term\,\cite{22-reza}. Authors of these works have used the free counterterms, which are for free cases with no nontrivial boundary conditions and obviously position-independent. It is of note that when non-trivial boundary condition or topology influences the quantum field, all elements of the renormalization program\,(\emph{e.g.}, the counterterms) should be consistent with it. Moreover, since counterterms are the responsible terms to renormalize the bare parameters in a problem, if they are not chosen properly, not all divergences may be removed correctly. It may cause some physical quantities, leading to a divergent value. By maintaining this idea, a systematic renormalization program was proposed by S.S. Gousheh et al., using which is facilitated the extraction of counterterms consistent with the boundary conditions\,\cite{SS.Gousheh.etal}. The obtained counterterms in their introduced renormalization program were position-dependent. All aspects about their procedures, involving the deduction of the counterterms from the $n$-point functions in the renormalized perturbation theory and its detailed calculations, have been reported previously. In the present study, using this procedure, we allowed the counterterms to be extracted automatically from the renormalization program, and by utilizing the obtained counterterms, we computed the vacuum energy of our system systematically up to the first order of coupling constant $\lambda$.
\begin{figure}[th] \hspace{0cm}\includegraphics[width=5cm]{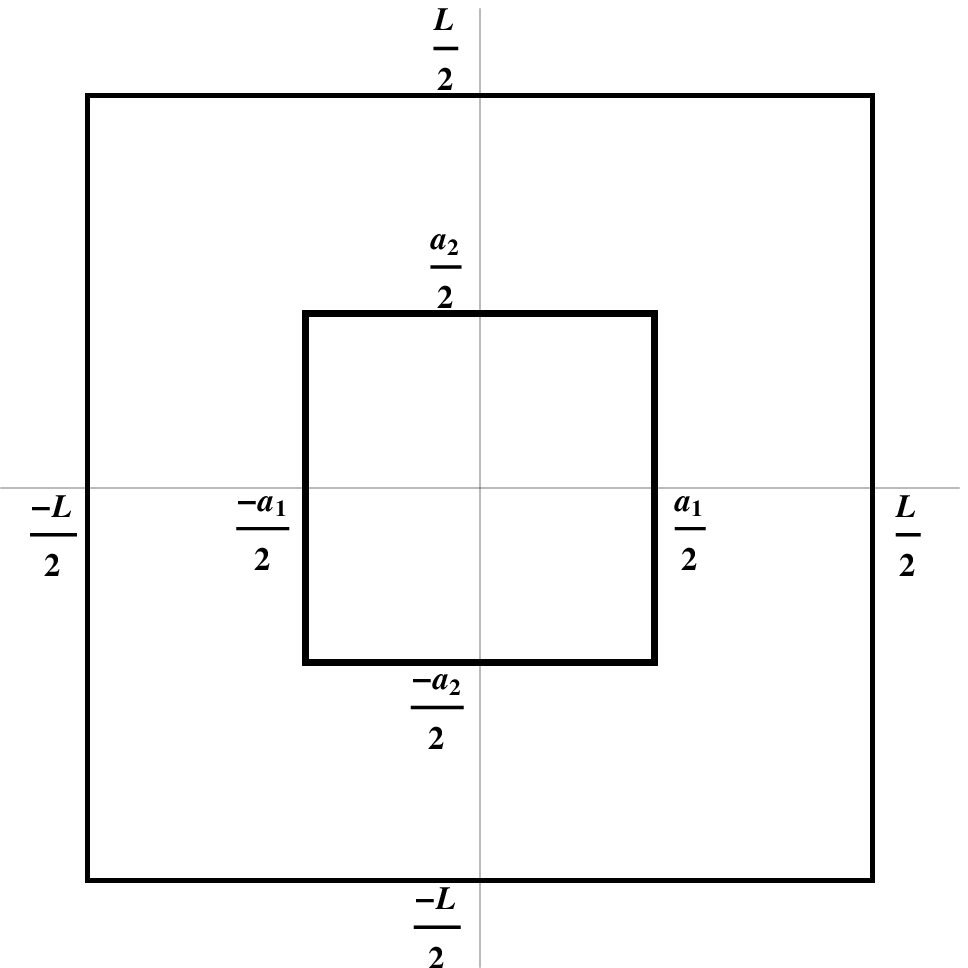}\hspace{0.5cm}\includegraphics[width=5cm]{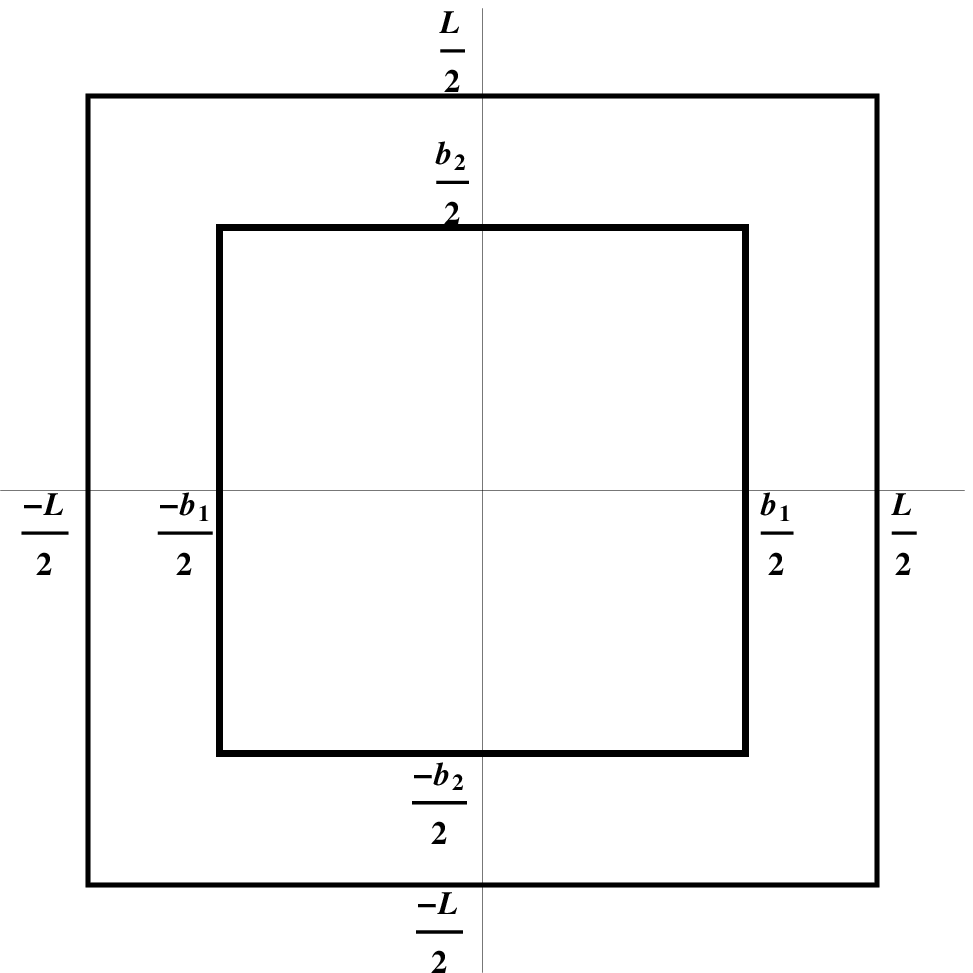}\caption{\label{fig.1} \small
  The Left figure is ``$A$ configuration" and the right one is ``$B$
  configuration".}
  \label{fig.1}
\end{figure}
\begin{figure}[th] \hspace{0cm}\includegraphics[width=5.5cm]{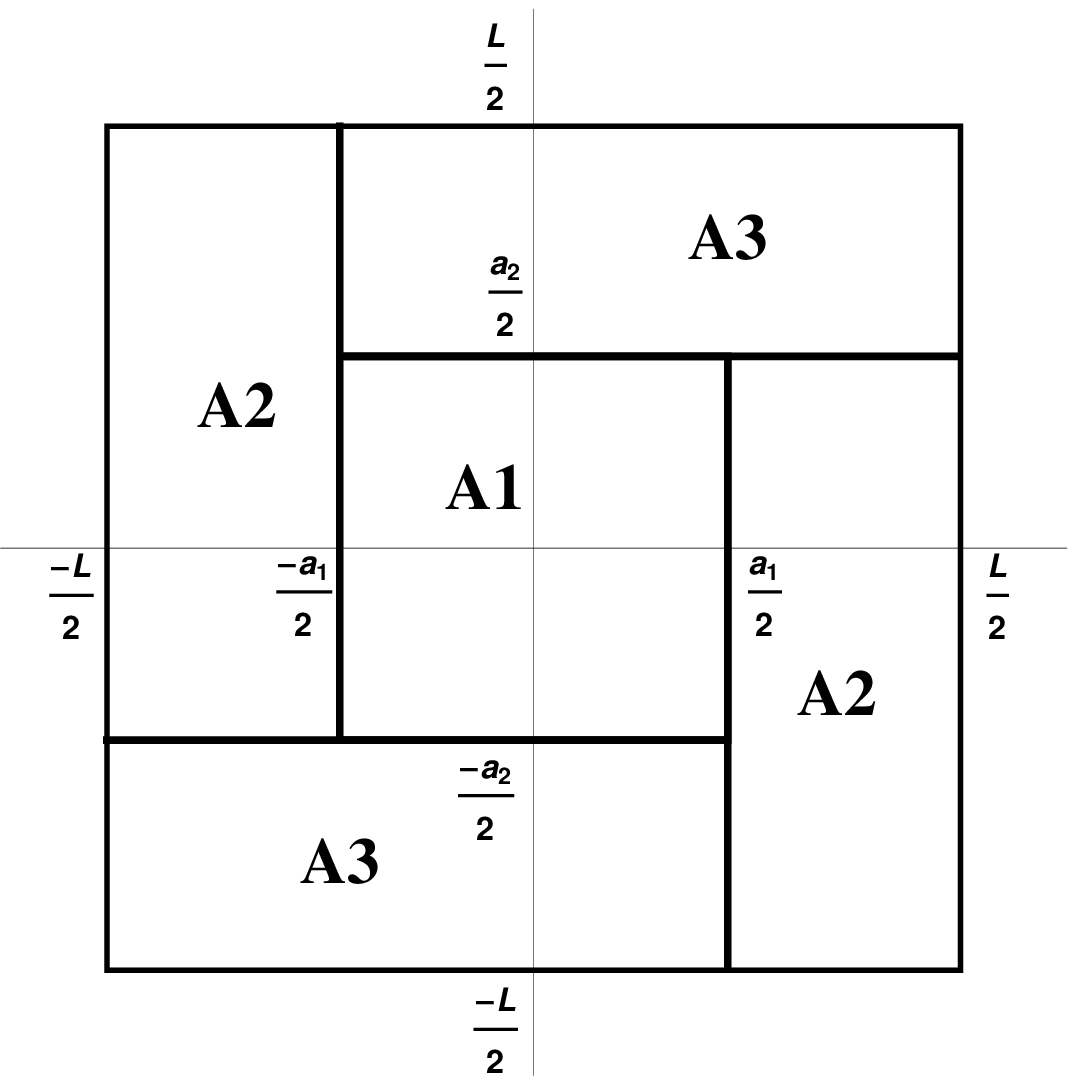}\hspace{0.5cm}\includegraphics[width=5.5cm]{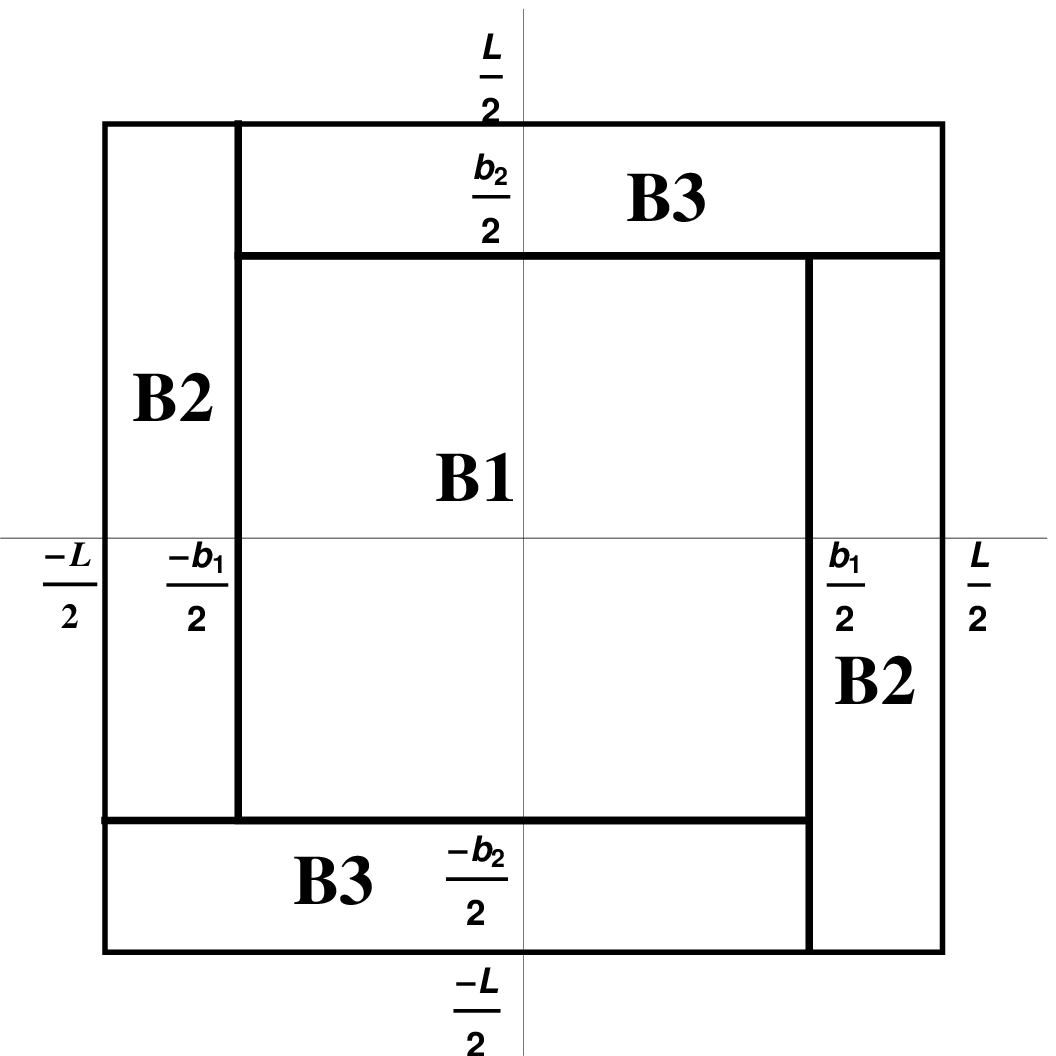}\caption{\label{fig.2} \small
  The Left figure is ``$A'$ configuration" and the right one is ``$B'$
  configuration". To calculate the Casimir energy, the zero-point energies of these two configurations should be subtracted according to Eq.\,\eqref{BSS.CAS.Def.}. In the final step, the size of configuration B goes to infinity, while the other parameters of the problem are maintained fixed.}
\label{fig.2}
\end{figure}
\par
The main task in the Casimir energy calculation is dealing with the infinite quantities. Therefore, the need to use a proper regularization technique was the main subject that physicists have consensus on it. In the literature, a wide range of topics in the Casimir energy has been proposed on a regularization technique and the advantage or disadvantage of them. Some of these known and important regularization techniques are the mode summation method\,\cite{mode.summation.}, Green's function technique\,\cite{Green.function.}, Schwinger's source theory\,\cite{schwinger.source.theory.}, or box subtraction scheme\cite{sphere.man}. In present work, we apply the \emph{Box Subtraction Scheme}\,(BSS), which is the slight modification of the Boyer's method\,\cite{boyer.}. In this method, two similar configurations are introduced and the vacuum energies of them are subtracted from each other. To reach a value for the Casimir energy of the original configuration, the size of the second configuration\,(\emph{i.e.}, the distance of two plates or radius of sphere) is considered as infinity. One of the main strength points in using of this regularization technique is the low necessity of resorting any analytic continuation. The imported parameters relevant to secondary configuration play the role of regulators in the calculation procedures. These additional regulators in reducing the use of analytic continuation are very effective. Also, these new regulators facilitate employing other regularization methods, such as the \emph{Cutoff Regularization Technique}\,(CRT), in the computing process. Indeed, the BSS supplemented by CRT automatically helps the divergence removal procedure to be conducted with more clarity. The BSS has been successful in presenting a physical answer for the Casimir energy problem designed in even spatial dimensions, which usually involves a high difficulty\,\cite{2dim.valuyan}. This scheme was also used for the calculation of the Casimir energy on a curved manifold and its result was consistent with known physical basis\,\cite{curved.valuyan}. Moreover, BSS has been successfully implemented as a regularization technique supplementing by the aforementioned renormalization program in the calculation of higher order radiative correction to the Casimir energy\,\cite{SS.Gousheh.etal}. In the present work, using the BSS, the Casimir energy is computed for the massive scalar field with Dirichlet boundary condition in a rectangle with cross-sectional area $a_1\times a_2$. To implement the BSS, two similar rectangles were introduced\,(Fig.\,(\ref{fig.1})). This figure shows a rectangle trapped in a larger rectangle\,(\emph{e. g.}, a square with a cross-section $L\times L$). The Casimir energy can now be defined as:
\begin{equation}\label{BSS.CAS.Def.}
  E_{{\rm{Cas.}}}  = \mathop {\lim }\limits_{b_1 /a,b_2 /a \to
  \infty } \left[ {\mathop {\lim }\limits_{L/b \to \infty } \left(
  {E_{A}- E_{B} } \right)} \right],
\end{equation}
where $E_A$ ($E_B$) is the vacuum energy of configuration A (B), $a\equiv \mbox{Max}\{a_{1},a_{2}\}$, and $b\equiv\mbox{Max}\{b_{1},b_{2}\}$. To compute the Casimir energy, it is necessary to have the vacuum energies in the whole configuration. However, the calculation for this quantity for the middle region of defined configurations in Fig.\,(\ref{fig.1}) is cumbersome. Therefore, to simplify the task, an alternative set of configurations was defined in Fig.\,(\ref{fig.2}) and we can then define the Casimir energy as in Eq.\,\eqref{BSS.CAS.Def.}, but with following replacements $A\rightarrow A'$ and $B\rightarrow B'$. Also, for the subtraction of the vacuum energies we have:
\begin{equation}\label{subtraction.vacuum.energy.}
  E_{A'}-E_{B'} = \big(E_{A1}+2E_{A2}+2E_{A3}\big)-\big( E_{B1}+2E_{B2}+2E_{B3}\big),
\end{equation}
where $E_{A1}$, $E_{A2}$,...,$E_{B3}$ is the vacuum energy of each region, separately. This new set of configuration was employed previously for the calculation of the Casimir energy in three spatial dimensions for an infinite rectangular waveguide\,\cite{waveguide.man}. The authors have proved that the additional lines in the middle region will not affect the Casimir energy of the original system and their obtained results will satisfy all necessary physical grounds. In the next section, to present accurate details for the BSS by utilizing this new set of configuration, we first calculated the leading order of Casimir energy for the free massive and massless scalar field in a rectangle. This quantity in a rectangular box with $p$ confined sides in $D$ spatial dimensions has already been calculated\,\cite{wolfram.locuz} using several analytic continuation techniques. Our method is free of any use of analytic continuation techniques and the answers are consistent with the previously reported results. In Section 3, using the BSS supplementing by the aforementioned renormalization program, we calculated the first-order radiative correction to the Casimir energy for a real massive scalar field in $\phi^4$ theory in a two-dimensional rectangular box\,(\emph{i.e.} a rectangle). In the following, by estimating the result in the massless case, all appropriate limits of the obtained answers were discussed. Finally, in Section 4, all physical aspects of using the applied methods and obtained results are summarized.

\section{Zero-Order Casimir Energy}
\label{sec:Leading.Cas.Cal.}
In two spatial dimensions, the total vacuum energy for the free massive scalar field confined with Dirichlet boundary condition in a rectangle with the cross-sectional area $a_1\times a_2$ is:
\begin{eqnarray}\label{Zero.Point.Energy.}
  E^{(0)} &=& \frac{1}{2}\sum_{n_1=1}^{\infty}\sum_{n_2=1}^{\infty}\omega_{\mathbf{k}},
\end{eqnarray}
where $\omega_{\mathbf{k}}=\sqrt{\big(\frac{n_1\pi}{a_1}\big)^2+\big(\frac{n_2\pi}{a_2}\big)^2+m^2}$ is the wave number.  To get the Casimir energy, as shown in Fig.\,(\ref{fig.2}) and Eq.\eqref{subtraction.vacuum.energy.}, the vacuum energy for each region should be calculated and the sum over vacuum energies of the whole configuration $B'$ should be subtracted from the configuration $A'$. The presentation of calculations for all regions is a tedious task, while the expression for the vacuum energy of each region is similar to others, with the only difference being the size of the region. Therefore, to simplify the task, we present the calculation for the original region $A1$ and spare presenting the details of calculations for the other regions. To do so, in each step of the calculation in Eq.\eqref{subtraction.vacuum.energy.}, we have only reported the first term in the right-hand side\,(rhs) of this equation\,(\emph{i.e.}, vacuum energy of original region $A1$) and ignored reporting of the other terms in the rhs of Eq.\,\eqref{subtraction.vacuum.energy.}.
\par
\begin{figure}[th] \hspace{0cm}\includegraphics[width=9.5cm]{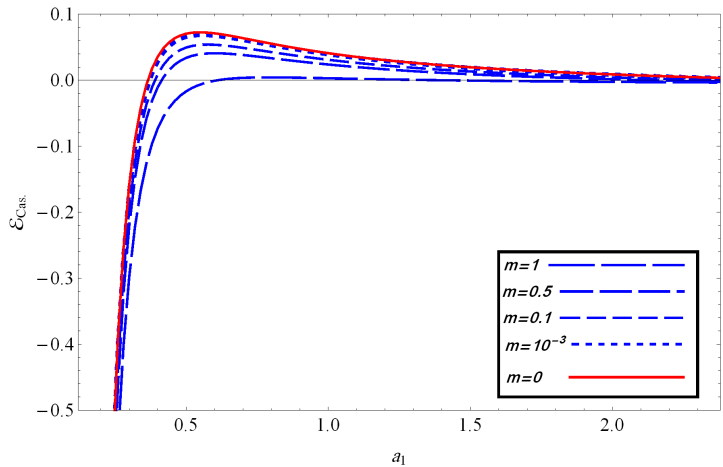}\hspace{0.5cm}
\caption{\label{fig.ZO.massiveMassless} \small
  The leading terms for the Casimir energy of massive and massless scalar fields in a rectangle with the cross-sectional area $a_1\times a_2$ as a function of $a_1$ for $\lambda=0.1$; the figure shows that the sequence of plots for the massive cases converges rapidly to the massless case and there is an insignificant difference between the figures of the massive cases for $m<0.001$, and the massless case.}
  \label{geometry}
\end{figure}

High modes render the sum in Eq.\eqref{Zero.Point.Energy.} formally divergent. To regularize the divergences by applying the following form of Abel-Plana Summation Formula\,(APSF), we convert all summation forms into the integral form\,\cite{Generalized.Abel.Plana.Saharian}:
\begin{equation}\label{APSF.1}
  \sum_{n=1}^{\infty}f(n)=-\frac{1}{2}f(0)+\int_{0}^{\infty}f(z)dz+i\int_{0}^{\infty}\frac{f(it)-f(-it)}{e^{2\pi t}-1}dt.
\end{equation}
Before starting the regularization procedure for Eq.\,\eqref{Zero.Point.Energy.}, by relation $\sum f(x,y) = \sum\frac{1}{2}\big(f(x,y)+f(y,x)\big)$, its expression was symmetrized in double argument $a_1$ and $a_2$. Now, using the APSF and our BSS introduced in Eqs.\,(\ref{BSS.CAS.Def.},\ref{subtraction.vacuum.energy.}, and \ref{APSF.1}), we have:
\begin{eqnarray}\label{applying.APSF.1}
  E^{(0)}_{A1}= \frac{1}{4}\sum_{n_1=1}^{\infty}\Bigg[\frac{-1}{2}\sqrt{\big(\frac{n_1\pi}{a_1}\big)^2+m^2}
  +\int_{0}^{\infty}\sqrt{\big(\frac{n_1\pi}{a_1}\big)^2+\big(\frac{x\pi}{a_2}\big)^2+m^2}dx+B_1(a_1,a_2)\Bigg]+\{a_1\leftrightarrow a_2\},
\end{eqnarray}
where $B_1(x,y)$ is the \emph{Branch-cut} term of APSF, which is:
 \begin{eqnarray}\label{Branch.cut.zero.order.1}
 B_1(x,y)=\frac{-2m^2y}{\pi}\int_{\sqrt{\mathcal{K}^2+1}}^{\infty}\frac{\sqrt{\eta^2+\mathcal{K}^2+1}}{e^{2my\eta}-1}d\eta
 =\frac{-1}{\pi}\sum_{j=1}^{\infty}\frac{\sqrt{\mathcal{K}^2+m^2}K_1(2yj\sqrt{\mathcal{K}^2+m^2})}{j},
\end{eqnarray}
where $\mathcal{K}=\frac{n_1\pi}{x}$ and the function $K_1(\alpha)$ is the modified Bessel function. The first and second terms in the bracket of Eq.\,\eqref{applying.APSF.1} are still divergent. To convert these summations into the integral form, by re-applying the APSF, we obtain:
\begin{eqnarray}\label{applying.APSF.2}
  E^{(0)}_{A1} = \frac{1}{4}\Bigg[\frac{1}{4}m-\underbrace{\frac{a_1+a_2}{2\pi}}_{\mathcal{C}_{A1}}\underbrace{\int_{0}^{\infty}\sqrt{\xi^2+m^2}d\xi}_{\mathcal{I}_\infty}
  +B_2(a_1)\hspace{8cm}\nonumber\\
  +\frac{a_1a_2}{\pi}\int_{0}^{\infty}\int_{0}^{\infty}\sqrt{\xi^2+\eta^2+m^2}d\xi d\eta+B_3(a_1,a_2)+\sum_{n_1=1}^{\infty} B_1(a_1,a_2)\Bigg]+\{a_1\leftrightarrow a_2\},\hspace{1.7cm}
\end{eqnarray}
where $B_2(x)$ and $B_3(x,y)$ are the Branch-cut terms of APSF and their values are:
\begin{eqnarray}\label{Branch.cut.zero.order.2}
B_2(x)&=&\frac{m^2x}{\pi}\int_{1}^{\infty}\frac{\sqrt{\eta^2-1}}{e^{2mx\eta}-1}d\eta=\frac{m}{2\pi}\sum_{j=1}^{\infty}\frac{K_1(2mxj)}{j},\nonumber\\
B_3(x,y)&=&\frac{-m^3xy}{2\pi}\int_{1}^{\infty}\frac{\eta^2-1}{e^{2mx\eta}-1}d\eta=\frac{-y}{8\pi x^2}\sum_{j=1}^{\infty}\frac{(2mxj+1)e^{-2mxj}}{j^3}.
\end{eqnarray}
The integral denoted by $\mathcal{I}_\infty$ in Eq.\eqref{applying.APSF.2} is divergent. To remove its infinity, we used the BSS and CRT. In this connection, a proper caution is needed when handling these infinity expressions. To do so, we first replace the upper limit of the integral $\mathcal{I}$ by a cutoff $\Lambda$. This replacement should be done for its counterparts in the other regions of configuration. As follows, a proper adjusting for $\Lambda$ and $\Lambda'$ with utilizing of Eq.\,\eqref{subtraction.vacuum.energy.} help all infinities due to integrals $\mathcal{I}$s be removed. Therefore we will have:
\begin{equation}\label{expand.first.integral.}
  \Big[\mathcal{C}_{A1}+\mathcal{C}_{A2}+\mathcal{C}_{A3}\Big]\mathcal{I}_\Lambda
  -\Big[\mathcal{C}_{B1}+\mathcal{C}_{B2}+\mathcal{C}_{B3}\Big]\mathcal{I}_{\Lambda'}=0.
  \end{equation}
Also, using Eq.\,\eqref{subtraction.vacuum.energy.}, there is no remained contribution from the second integral term of Eq.\,\eqref{applying.APSF.2} as:
\begin{equation}\label{removing of integral term.BSS.}
  \Bigg(a_1a_2+2\frac{L-a_1}{2}\frac{L+a_2}{2}+2\frac{L+a_1}{2}\frac{L-a_2}{2}-\{a_i \leftrightarrow b_i\}\Bigg)
  \int_{0}^{\infty}\int_{0}^{\infty}\sqrt{\xi^2+\eta^2+m^2}d\xi d\eta=0
\end{equation}
Therefore, the only remaining finite terms from Eq.\,\eqref{applying.APSF.2} are the Branch-cut terms $B_1(x,y)$, $B_2(x)$ and $B_3(x,y)$. At the last step, we computed the limits displayed in Eq.\,\eqref{BSS.CAS.Def.}. After calculating these limits, the expression for the leading order Casimir energy of massive scalar field confined in a rectangle becomes:
\begin{eqnarray}\label{Leading.Order.Massive.Casimir.}
  E^{(0)}_{\textrm{Cas.} }= \frac{1}{4}\Big[B_2(a_1)+B_3(a_1,a_2)+\sum_{n=1}^{\infty}B_1(a_1,a_2)+\{a_1\leftrightarrow a_2\}\Big]
\end{eqnarray}
The main particular limit for the Casimir energy is usually known as the massless limit. After this limit, Eq.\,\eqref{Leading.Order.Massive.Casimir.} is converted to:
\begin{eqnarray}\label{Leading.Order.Massless.Casimir.}
  E^{(0)}_{\textrm{Cas.}} \buildrel {m\to 0}\over\longrightarrow \frac{1}{4}\Bigg[\frac{\pi}{24a_1}-\frac{a_2}{8\pi a_1^2}\zeta(3)-\sum_{n,j=1}^{\infty}\frac{n K_1(2\pi nja_2/a_1)}{ja_1}+\{a_1\leftrightarrow a_2\}\Bigg]
\end{eqnarray}
Results obtained for the Casimir energy in both massive and massless cases are obviously finite and consistent with previously reported results\,\cite{wolfram.locuz}. In Fig.\,(\ref{fig.ZO.massiveMassless}), the zero order of the Casimir energy density as a function of one side of the rectangle is plotted. In this figure, we show the sequence of plots for $m = \{1, 0.1, 0.01, 0.001, 0\}$. Apparently, this figure shows the sequence of plots for the massive cases converges rapidly to the massless ones when $m\to 0$.

\section{First-Order Radiative Correction to the Casimir Energy}
\label{sec:RC.Cas.Cal.}
In this section, we calculate the first order radiative correction to the Casimir energy for the real massive scalar field in $\lambda\phi^4$ theory in a two-dimensional rectangular box\,(\emph{i.e.} a rectangle). In the presence of the nontrivial boundary conditions, as outlined in the Introduction, the counterterms are naturally obtained position-dependent and their computation is usually conducted from the appropriate  $n$-point functions. The renormalization procedure, the deduction of the counterterms, and the final general form of the first order correction to the total vacuum energy were completely discussed in Refs.\,\cite{SS.Gousheh.etal}. Therefore, in the preset work we use only their conclusions:
\begin{equation}\label{VacuumEn.firstorder.}
  E^{(1)}=\frac{-\lambda}{8} \int_{S} G^2 (x,x)d^2\mathbf{x},
\end{equation}
where $G(x, x')$ is the propagator of the real scalar field. The final expression for the Green's function of the massive scalar field with Dirichlet boundary condition in the rectangle with cross-sectional area $a_1\times a_2$, after Wick rotation, can be written as:
\begin{eqnarray}\label{Green.function}
  G(x,x')=\frac{2}{a_1a_2}
   \sum_{n_1,n_2=1}^{\infty}\frac{1}{\omega_{\mathbf{k}}}\sin\Big(\frac{n_1\pi}{a_1}(x+\frac{a_1}{2})\Big)\sin\Big(\frac{n_2\pi}{a_2}(y+\frac{a_2}{2})\Big)
   \sin\Big(\frac{n_1\pi}{a_1}(x'+\frac{a_1}{2})\Big) \sin\Big(\frac{n_2\pi}{a_2}(y'+\frac{a_2}{2})\Big),
\end{eqnarray}
where $\omega^2_{\mathbf{k}}=(\frac{n_1\pi}{a_1})^2+(\frac{n_2\pi}{a_2})^2+m^2$ and $m$ is the mass of the field. It is of note that to find the Casimir energy according to Fig.\,(\ref{fig.2}) and Eq.\,\eqref{subtraction.vacuum.energy.}, the total vacuum energies of two configurations should be subtracted from each other. Therefore, the vacuum energy expression for all regions should be separately available. However, the presentation of the details of calculation for all regions is a time-consuming process, while the expression for the vacuum energy of each region is similar to others, being different only in the size of the region; therefore, to simplify the task, we only track and report the details of calculation for one region\,(\emph{i.e.}, the original region $A1$). In the following, the subtraction defined in Eq.\eqref{subtraction.vacuum.energy.} was calculated. However, in each step, we have only tracked and reported the first term in the rhs of Eq.\,\eqref{subtraction.vacuum.energy.}\,(vacuum energy of original region $A1$) and reporting of the other terms in the rhs of Eq.\,\eqref{subtraction.vacuum.energy.} were ignored. Thus, for the total vacuum energy of region $A1$ displayed in Fig.\,(\ref{fig.2}), after substituting the Green's function expression given in Eq.\,\eqref{Green.function} with Eq.\,\eqref{VacuumEn.firstorder.} and calculating all integrals over space, we have:
\begin{eqnarray}\label{All.Four.Terms}
  E^{(1)}_{A1}=\frac{-\lambda L}{32a_1a_2}\sum_{n_1,n'_1=1}^{\infty}\sum_{n_2,n'_2=1}^{\infty}\frac{1}{\omega_{\mathbf{k}}\omega_{\mathbf{k'}}}
  \bigg[1+\frac{1}{2}\delta_{n_1,{n'}_1}+\frac{1}{2}\delta_{n_2,{n'}_2}+\frac{1}{4}\delta_{n_1,{n'}_1}\delta_{n_2,{n'}_2}\bigg],
\end{eqnarray}
where $\omega^2_{\mathbf{k'}}=(\frac{n'_1\pi}{a_1})^2+(\frac{n'_2\pi}{a_2})^2+m^2$. Obviously, for high frequency all summations in Eq.\,\eqref{All.Four.Terms} are divergent. To regularize the divergences and remove their infinities via BSS, these summation forms should are first converted into the integral form. To do so, we have used again the APSF defined in Eq.\,\eqref{APSF.1}. However, applying this formula even for the vacuum energy of region $A1$ is still lengthy. So, for the sake of transparency in presenting of the calculation, we split the bracket of Eq.\,\eqref{All.Four.Terms} into four parts. In the following sub-sections, the calculation for each term in the bracket of Eq.\,\eqref{All.Four.Terms} was conducted. Finally, the sum of all remaining finite pieces from these four parts will be discussed.

\subsubsection{The First Term}
\label{sub.section.first.term}
Before applying the APSF, using $\sum_{i,j} f(x,y) = \sum_{i,j}\frac{1}{2}\big(f(x,y)+f(y,x)\big)$, we first symmetrized all expressions in Eq.\,\eqref{All.Four.Terms} with respect to the double arguments $a_1$ and $a_2$. Then, for the first term of Eq.\,\eqref{All.Four.Terms} by applying the APSF on summation over $n_1$ we have:
\begin{eqnarray}\label{Term1.A1}
  T_1&=&\frac{-\lambda}{32a_1a_2}\bigg[\frac{1}{2}\sum_{n_1,n_2=1}^{\infty}\frac{1}{\omega_{\mathbf{k}}}+\{a_1\leftrightarrow a_2\}\bigg]^2\hspace{8.5cm}\nonumber \\
  &=& \frac{-\lambda}{32a_1a_2}\bigg[\frac{1}{2}\sum_{n_2=1}^{\infty}\bigg(\frac{-1}{2}\frac{1}{\sqrt{(\frac{n_2\pi}{a_2})^2+m^2}}
  +\int_{0}^{\infty}\frac{dx}{\sqrt{(\frac{x\pi}{a_1})^2+(\frac{n_2\pi}{a_2})^2+m^2}}+B_1(m;a_1,a_2)\bigg)+\{a_1\leftrightarrow a_2\}\bigg]^2,
\end{eqnarray}
where $B_1(m;x,y)=\frac{2x}{\pi}\sum_{j=1}^{\infty}K_0(2xj\sqrt{(\frac{n_2\pi}{y})^2+m^2})$ is the Branch-cut term of APSF and its value is finite. Also $K_0(\alpha)$ is the modified Bessel function. Re-applying the APSF on the remaining divergent terms of Eq.\,\eqref{Term1.A1} gives:
\begin{eqnarray}\label{Term1.A2}
  T_1=\frac{-\lambda}{32a_1a_2}\Bigg[\frac{1}{2}\Bigg(\frac{1}{4m}-\frac{1}{2}\underbrace{\int_{0}^{\infty}\frac{dx}{\sqrt{(\frac{x\pi}{a_2})^2+m^2}}}_{I_1(a_2)}
  +B_2(m;a_2)+\underbrace{\int_{0}^{\infty}\int_{0}^{\infty}\frac{dxdy}{\sqrt{(\frac{x\pi}{a_1})^2+(\frac{y\pi}{a_2})^2+m^2}}}_{I_2(a_1,a_2)}
  \hspace{2cm}\nonumber\\
  -\frac{1}{2}\underbrace{\int_{0}^{\infty}\frac{dx}{\sqrt{(\frac{x\pi}{a_1})^2+m^2}}}_{I_1(a_1)}
  +B_3(m;a_1,a_2)+\sum_{n_2=1}^{\infty}B_1(m;a_1,a_2)\Bigg)+\{a_1\leftrightarrow a_2\}\Bigg]^2,\hspace{2cm}
\end{eqnarray}
where $B_2(m;x)$ and $B_3(m;x,y)$ are the Branch-cut term of APSF and their values are:
\begin{eqnarray}\label{Branchcut.term.first.term.}
 B_2(m;x)=\frac{-x}{\pi}\sum_{j=1}^{\infty}K_0(2mxj),\hspace{2cm}
 B_3(m;x,y)= \frac{-x}{2\pi}\ln\big(1-e^{-2my}\big).
\end{eqnarray}
The integral terms shown by $I_1(x)$ and $I_2(x,y)$ in Eq.\,\eqref{Term1.A2} have still a divergent value. To remove their infinities, similar for the case of Eqs.\,(\ref{expand.first.integral.} and \ref{removing of integral term.BSS.}), the CRT and BSS should be implemented. Therefore, we replace the upper limits of integrals with multiple cutoffs. Then, by calculating integrations and expanding the results in the infinite limit of cutoffs, the divergent part of each result is manifested. Same as this scenario should be conducted for the counterpart terms in the other regions. Now, it can be shown that by adjusting a proper value for cutoffs and using the BSS (Eq.\,\eqref{subtraction.vacuum.energy.}) all divergent parts of these integrals in Eq.\,\eqref{Term1.A2} would be removed. The remaining finite parts related to each integral become:
\begin{eqnarray}\label{Term1.Is.}
  I_1(x)\longrightarrow \frac{x}{\pi}\ln2,\hspace{2cm}I_{2}(x,y)\longrightarrow 0.
\end{eqnarray}
Therefore, the final expression $T_1$ for region $A1$ becomes:
\begin{eqnarray}\label{first.term.final.form.}
   T_1=\frac{-\lambda}{32a_1a_2}\Bigg[\frac{1}{2}\Bigg(\frac{1}{4m}-\frac{a_1+a_2}{2\pi}\ln2
  +B_2(m;a_2)+B_3(m;a_1,a_2)+\sum_{n_2=1}^{\infty}B_1(m;a_1,a_2)\Bigg)+\{a_1\leftrightarrow a_2\}\Bigg]^2,
\end{eqnarray}
\subsubsection{The Second and Third Term}
The second and third terms of Eq.\,\eqref{All.Four.Terms} are symmetric relative to the displacement of $a_1$ and $a_2$. Therefore, after applying the APSF, the summation of $T_2+T_3$ can be written as:
\begin{eqnarray}\label{Term2.A1.}
  T_2+T_3&=&\frac{-\lambda}{32a_1a_2}\frac{1}{2}\sum_{n_1=1}^{\infty}\sum_{n_2,n'_2=1}^{\infty}\frac{1}{\omega_{\mathbf{k}}}\frac{1}{\omega_{\mathbf{k'}}}
  \hspace{8cm}\nonumber \\&=& \frac{-\lambda}{32a_1a_2}\frac{1}{2}\sum_{n_1=1}^{\infty}\bigg[\underbrace{\frac{-1}{2}\frac{1}{\sqrt{(\frac{n_1\pi}{a_1})^2+m^2}}
  +\int_{0}^{\infty}\frac{dx}{\sqrt{(\frac{n_1\pi}{a_1})^2+(\frac{x\pi}{a_2})^2+m^2}}}_{\mathcal{A}}+B_1(m;a_2,a_1)\bigg]^2+\{a_1\leftrightarrow a_2\}\nonumber\\
  &=&\frac{-\lambda}{32a_1a_2}\frac{1}{2}\sum_{n_1=1}^{\infty}\Big[\mathcal{A}^2+2\mathcal{A}B_1(m;a_2,a_1)+B_1^2(m;a_2,a_1)\Big]+\{a_1\leftrightarrow a_2\}.
\end{eqnarray}
At the next step, the APSF was re-applied on the first term in the rhs of Eq.\,\eqref{Term2.A1.}. It has to be noted that values of all new created Branch-cut terms after this applying of APSF is zero and no contribution of them remains in $T_2+T_3$. Thus:
\begin{eqnarray}\label{Term2.A2.}
  \small T_2+T_3=\frac{-\lambda}{32a_1a_2}\frac{1}{2}\Bigg[\frac{-1}{8m^2}+\frac{1}{4}\underbrace{\int_{0}^{\infty}\frac{dx}{\frac{x^2\pi^2}{a_1^2}+m^2}}_
  {=\frac{a_1}{2m}}+\frac{1}{2m}\underbrace{\int_{0}^{\infty}\frac{dx}{\sqrt{\frac{x^2\pi^2}{a_2^2}+m^2}}}_{I_1(a_2)}\hspace{6.6cm}\nonumber\\
 {\small -\underbrace{\int_{0}^{\infty}\frac{dy}{\sqrt{\frac{y^2\pi^2}{a_1^2}+m^2}}\int_{0}^{\infty}\frac{dx}{\sqrt{\frac{y^2\pi^2}{a_1^2}
  +\frac{x^2\pi^2}{a_2^2}+m^2}}}_{I_{3}(a_1,a_2)}-\frac{1}{2}\bigg(\underbrace{\int_{0}^{\infty}\frac{dx}{\sqrt{\frac{x^2\pi^2}{a_2^2}+m^2}}}_{I_1(a_2)}\bigg)^2
 +\underbrace{\int_{0}^{\infty}dy\bigg(\int_{0}^{\infty}\frac{dx}{\sqrt{\frac{y^2\pi^2}{a_1^2}+\frac{x^2\pi^2}{a_2^2}+m^2}}\bigg)^2}_{I_{4}(a_1,a_2)}}
  \hspace{0.1cm}\nonumber\\
 \small +\sum_{n_1=1}^{\infty}\Bigg(2B_1(m;a_2,a_1)
  \underbrace{\int_{0}^{\infty}\frac{dx}{\sqrt{\frac{n_1^2\pi^2}{a_1^2}+\frac{x^2\pi^2}{a_2^2}+m^2}}}_{I_{5}(a_1,a_2)}\Bigg)
  +\sum_{n_1=1}^{\infty}B^2_1(m;a_2,a_1)\Bigg]+\{a_1 \leftrightarrow a_2\}.\hspace{3.2cm}
\end{eqnarray}
The terms $I_{3}(x,y)$, $I_{4}(x,y)$, and $I_{5}(x,y)$ have a divergent value and thus must be properly regularized. To do so, same as what happened for Eq.\,\eqref{Term1.A2}, we prefer to use the CRT and BSS again. Therefore, the remaining finite parts for each term will be obtained as:
\begin{equation}\label{Is. Remaining}
  I_{3}(x,y)\longrightarrow \frac{xy}{\pi^2}\ln^22,\hspace{2cm}  I_{4}(x,y)\longrightarrow 0, \hspace{2cm}  I_{5}(x,y)\longrightarrow \frac{y}{\pi}\ln2.
\end{equation}
Using Eqs.\,(\ref{Term1.Is.},\ref{Term2.A2.},\ref{Is. Remaining}), the final expression for $T_2+T_3$ becomes:
\begin{eqnarray}\label{Term2&3.final}
  T_2+T_3=\frac{-\lambda}{32a_1a_2}\frac{1}{2}\Bigg[\frac{-1}{8m^2}+\frac{a_1}{8m}+\frac{a_2}{2\pi m}\ln2-\frac{a_1a_2}{\pi^2}\ln^22
  -\sum_{n_1=1}^{\infty}\frac{B_{1}(m;a_2,a_1)}{\sqrt{\frac{n_1^2\pi^2}{a_1^2}+m^2}}\hspace{4.8cm}\nonumber\\-\frac{a_2^2}{2\pi^2}\ln^22+
  \frac{2a_2}{\pi}\ln2\sum_{n_1=1}^{\infty}B_{1}(m;a_2,a_1)
  +\sum_{n_1=1}^{\infty}B^2_{1}(m;a_2,a_1)\Bigg]+\{a_1 \leftrightarrow a_2\},\hspace{2cm}
\end{eqnarray}

\subsubsection{The Fourth Term}
For the last term of Eq.\eqref{All.Four.Terms}, after applying of APSF on both summations, it can be easily shown that, the values of all new created Branch-cut terms are zero and thus no contribution of it remains in $T_4$. So,
\begin{eqnarray}\label{Term4.A1.}
  T_4&=&\frac{-\lambda}{32a_1a_2}\frac{1}{8}\sum_{n_1=1}^{\infty}\sum_{n_2=1}^{\infty}\frac{1}{\omega_{\mathbf{k}^2}}+\{a_1 \leftrightarrow a_2\}\nonumber\\
  &=&\frac{-\lambda}{32a_1a_2}\frac{1}{8}\sum_{n_2=1}^{\infty}\Bigg[\frac{-1}{2}\frac{1}{\frac{n_2^2\pi^2}{a_1^2}+m^2}
  +\int_{0}^{\infty}\frac{dx}{(\frac{x\pi}{a_1})^2+(\frac{n_2\pi }{a_2})^2+m^2}\Bigg]+\{a_1 \leftrightarrow a_2\} \\
  \tiny &=&\frac{-\lambda}{32a_1a_2}\frac{1}{8}\Bigg[\frac{1}{4m^2}
 -\frac{1}{2}\underbrace{\int_{0}^{\infty}\frac{dx}{(\frac{x\pi}{a_2})^2+m^2}}_{=\frac{a_2}{2m}}
  -\frac{1}{2}\underbrace{\int_{0}^{\infty}\frac{dx}{(\frac{x\pi}{a_1})^2+m^2}}_{=\frac{a_1}{2m}}
  +\underbrace{\int_{0}^{\infty}\int_{0}^{\infty}\frac{dxdy}{(\frac{x\pi}{a_1})^2+(\frac{y\pi }{a_2})^2+m^2}}_{I_6(a_1,a_2)}\Bigg]+\{a_1 \leftrightarrow a_2\},\nonumber
\end{eqnarray}
The integral $I_6(a_1,a_2)$  has a divergent value and by utilizing the CRT and BSS, the remaining finite expression for this integral becomes: $I_6(a_1,a_2)\rightarrow \frac{a_1a_2}{2\pi}\ln2$. The final remaining terms for $T_4$ will be obtained as:
\begin{equation}\label{Term4. final}
  T_4=\frac{-\lambda}{32a_1a_2}\frac{1}{4}\Bigg[\frac{1}{4m^2}-\frac{a_1+a_2}{4m}+\frac{a_1a_2}{2\pi}\ln2\Bigg].
\end{equation}
In above sub-sections, four parts of Eq.\,\eqref{All.Four.Terms} were discussed separately. Also, using Eq.\,\eqref{subtraction.vacuum.energy.} all infinite parts of them were removed by their counterparts in the other regions. By summing up all remaining expressions $T_1$, $T_2$, $T_3$, and $T_4$ according to Eq.\,\eqref{All.Four.Terms}, many cancellations occur internally. All remaining terms, at this step, are convergent for any finite values of $a_1$, $a_2$, $L$ and $m\neq0$. At final step, according to Eq.\,\eqref{BSS.CAS.Def.} the limit $L/b\to\infty$ and $b/a\to\infty$ should be calculated. After this limit, the contribution of all remaining terms related to all regions except for region $A1$ vanishes. Finally, the first order radiative correction to the Casimir energy for massive scalar field confined in a rectangle with the cross-sectional area $a_1\times a_2$ is obtained as:
\begin{eqnarray}\label{Final.RC.Massive.}
     E_{\mbox{\small{Cas.}}}^{(1)}=\frac{-\lambda}{32a_1 a_2}\bigg\{\frac{1}{4m}B_{2}(m;a_1)-\frac{a_1}{8\pi m}\ln(1-e^{-2m a_2})
     +\frac{1}{4m}\sum_{n=1}^{\infty}B_{1}(m;a_1,a_2)-\frac{a_1\ln2}{2\pi}B_{2}(m;a_1)\hspace{3cm}\nonumber\\
     +\frac{a_1^2\ln2}{4\pi^2}\ln(1-e^{-2ma_2})
     +\frac{a_1a_2\ln2}{4\pi^2}\ln(1-e^{-2ma_1})-\frac{a_1\ln2}{2\pi}\sum_{n=1}^{\infty}B_{1}(m;a_2,a_1)+\frac{1}{4}B_{2}^2(m;a_2)
     \hspace{1.2cm}\nonumber\\
     +\frac{1}{4}B_{2}(m;a_2)B_{2}(m;a_1)-\frac{a_1}{4\pi}\ln(1-e^{-2ma_2})\Big(B_{2}(m;a_2)+B_{2}(m;a_1)\Big)
      \hspace{4.6cm}\nonumber\\
     +\frac{1}{2}(B_{2}(m;a_2)+B_{2}(m;a_1))\sum_{n=1}^{\infty}B_{1}(m;a_1,a_2)
     +\frac{a_1a_2}{16\pi^2}\ln(1-e^{-2ma_1})\ln(1-e^{-2ma_2})\hspace{2.5cm}\nonumber\\
     -\frac{1}{4\pi}\Big(a_1\ln(1-e^{-2ma_2})+a_2\ln(1-e^{-2ma_1})\Big)\sum_{n=1}^{\infty}B_{1}(m;a_1,a_2)+\frac{a_2^2}{16\pi^2}\ln^2(1-e^{-2ma_1})
     \hspace{2.1cm}\nonumber\\
     +\frac{1}{4}\Big(\sum_{n=1}^{\infty}B_{1}(m;a_1,a_2)\Big)^2
     +\frac{1}{4}\Big(\sum_{n=1}^{\infty}B_{1}(m;a_1,a_2)\Big)\Big(\sum_{n=1}^{\infty}B_{1}(m;a_2,a_1)\Big)
     \hspace{4.8cm}\nonumber\\+\frac{1}{2}\sum_{n=1}^{\infty}\frac{B_{1}(m;a_1,a_2)}{\sqrt{\frac{n^2\pi^2}{a_2}+m^2}}
     +\frac{1}{2}\sum_{n=1}^{\infty}B^2_{1}(m;a_1,a_2)
     \Bigg\}+\{a_1\leftrightarrow a_2\},\hspace{6.4cm}
\end{eqnarray}
\begin{figure}[th] \hspace{0cm}\includegraphics[width=8.5cm]{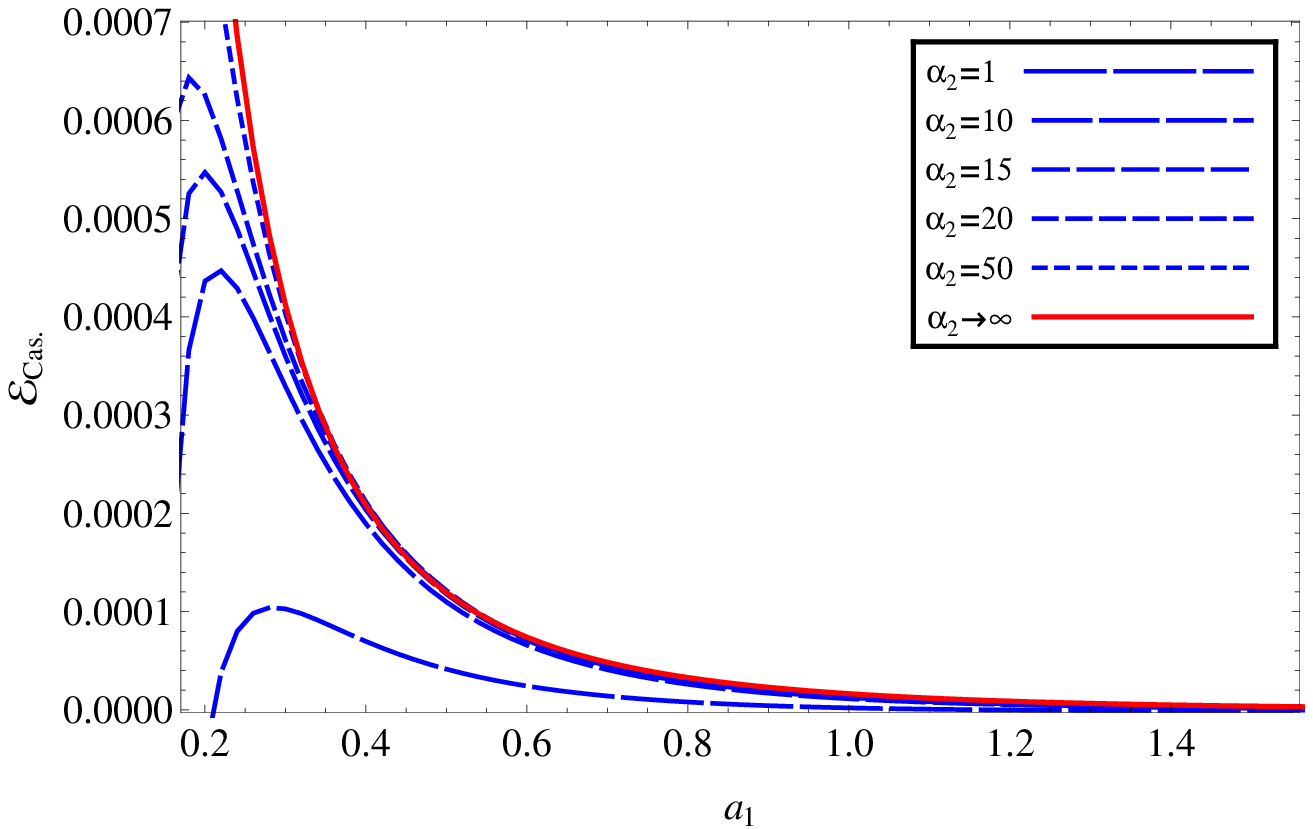}\hspace{0.5cm}
\caption{\label{fig.massive} \small
  The first order radiative correction to the Casimir energy density for the massive scalar field in a rectangle with the cross-sectional area $a_1\times a_2$ as a function of $a_1$ for the sequence values of $a_2=\{1,10,15,20,50\}$\,(the dashed lines); the solid line shows the radiative correction to the Casimir energy density between two parallel plates\,(to find its expression see Eq.\,(20) in Ref.\,\cite{2dim.valuyan}) as a function of the distance of the plates. This set of figure apparently shows when a given size of the rectangle\,(\emph{e.g.}, the size $a_2$) goes to infinity, the Casimir energy value approaches to the ones for two parallel plates. The values of the mass of the field and coupling constant in all plots are $m=1$ and $\lambda=0.1$, respectively.}
  \label{geometry}
\end{figure}
\begin{figure}[th] \hspace{0cm}\includegraphics[width=8.5cm]{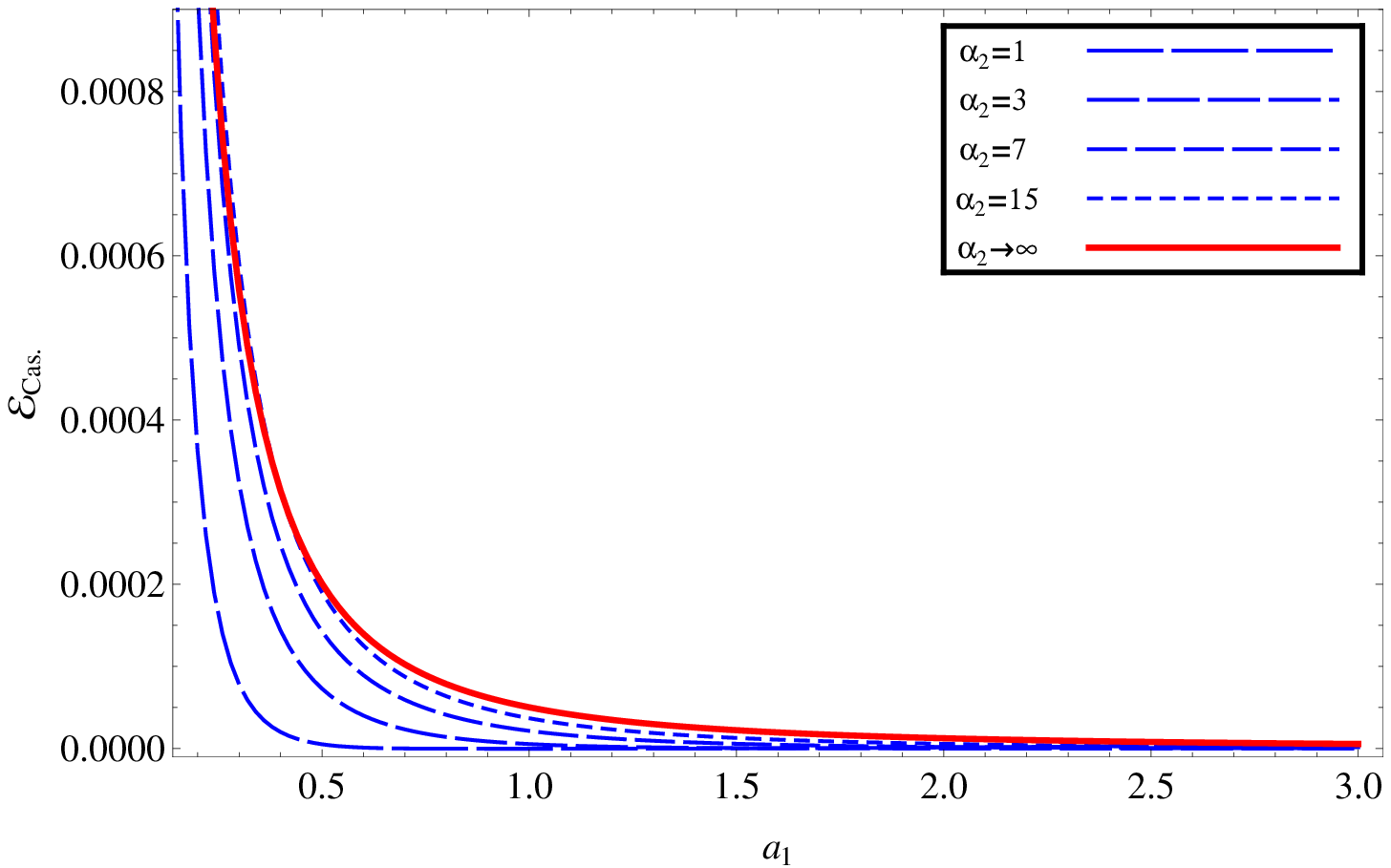}\hspace{0.5cm}
\caption{\label{fig.massless} \small
  The first order radiative correction to the Casimir energy density for the massless scalar field in a rectangle with the cross-sectional area $a_1\times a_2$ as a function of $a_1$ for the sequence values of $a_2=\{1,3,7,15\}$\,(the dashed lines); the solid line shows the radiative correction to the Casimir energy density between two parallel plates\,(to find its expression see Eq.\,(23) in Ref.\,\cite{2dim.valuyan}) as a function of the distance of the plates. This set of figure apparently shows when one size of the rectangle\,(\emph{e.g.}, the size $a_2$) goes to infinity, the Casimir energy value approaches to the ones for two parallel plates. The value of coupling constant in all plots is considered as $\lambda=0.1$.}
  \label{geometry}
\end{figure}
For any finite values of mass $m\neq0$ the above expression is finite and its computation should be partly conducted numerically. This result is also consistent with the previously reported result for two parallel plates in proper limits\,\cite{2dim.valuyan}. Fig.\,(\ref{fig.massive}) presents the Casimir energy density as a function of one side of the rectangle\,(\emph{e.g.}, the side $a_1$) for multiple values of $a_2=\{1,10,15,20,50\}$. This figure shows when the size of one side of the rectangle goes to infinity the Casimir energy value approaches to the ones for two parallel plates. This consequence is acceptable on physical grounds.
\par
An important extreme limit for the obtained quantity in Eq.\,\eqref{Final.RC.Massive.} is the massless limit. However, the direct calculation of the massless limit from Eq.\,\eqref{Final.RC.Massive.} is cumbersome and it leads to a divergent value. Hence, we go back to the vacuum energy expression, given in Eq.\,\eqref{All.Four.Terms}, and put the mass of the field as zero and start the calculation process again. The significant parts of the calculations are similar to what occurred in the problem for the massive case. Hence, we present the details of this computation in appendix\,\ref{Appendix.A} and only report the final output here. The final expression for the first order radiative correction to the Casimir energy for the massless scalar field is:
\begin{eqnarray}\label{Massless.Casimir.Expression.}
 E_{\mbox{\small{Cas.}}}^{(1)}\buildrel m=0\over{\longrightarrow}\frac{-\lambda}{32a_1a_2}\frac{1}{2}\sum_{n_1=1}^{\infty}\Bigg\{-\frac{a_1}{\pi n_1}B_{1}(0;a_2,a_1)+B^2_{1}(0;a_2,a_1)+\frac{2a_2\ln2}{\pi}B_{1}(0;a_2,a_1)\Bigg\}+\{a_1\leftrightarrow a_2\},
\end{eqnarray}
where $B_{1}(0;x,y)=\frac{2x}{\pi}\sum_{j=1}^{\infty}K_{0}\big(\frac{2\pi xjn}{y}\big)$ and $K_0(\alpha)$ is the modified Bessel function. To investigate the obtained result more precisely, we checked the limit of Eq.\,\eqref{Massless.Casimir.Expression.} when the one side of the rectangle goes to infinity. In this limit, it is expected that the Casimir energy value would be approached to the Casimir energy of two parallel plates. The first order radiative correction to the Dirichlet Casimir energy density for two parallel plates with distance $a_1$ for the massless scalar field, in $\phi^4$ theory, was reported as $\frac{\lambda}{128\pi^2a_1^2}(0.6349208)$\,\cite{2dim.valuyan}. In Fig.\,(\ref{fig.massless}), we show that when one side of the rectangle\,(\emph{e. g.}, the side $a_2$) goes to infinity, the Casimir energy given in Eq.\,\eqref{Massless.Casimir.Expression.} approaches to the one for two parallel plates. This is a consistent outcome and it also agrees with known physical basis.
\section{Conclusion}
In this paper, the zero and the first order radiative correction to the Casimir energy for the massive scalar field with Dirichlet boundary condition was computed in a two-dimensional rectangular box\,(i.e. a rectangle). This calculation for the first order radiative correction to the Casimir energy is new and our used method in the problem is also different with the usual and known  method in the literature. To renormalize the bare parameters of the theory, we allowed the counterterm be position-dependent. Such counterterms allow all influences from the boundary conditions be imported in the renormalization program. The deduction of these counterterms was conducted by a systematic perturbation theory in a few previous works and in this study, maintaining that idea, we used only their conclusions. Another noteworthy point in this article is the use of special regularization technique, which we have named it the Box Subtraction Scheme\,(BSS). In this regularization procedure, two similar configurations are defined and the vacuum energies of them are subtracted from each other. This subtraction procedure supplemented by the cutoff regularization technique helps to reduce the need to use any analytic continuation technique. All obtained results are consistent with previously reported results in the appropriate limits.
\appendix
\section{Calculations for the Massless Case}
\label{Appendix.A}
To compute the radiative correction to the Casimir energy for the massless scalar field in the rectangle we go back to Eq.\,\eqref{All.Four.Terms} and put the mass of the field as zero. At the next step, for the first term of Eq.\,\eqref{All.Four.Terms}, using $\sum_{i,j} f(x,y) = \sum_{i,j}\frac{1}{2}\big(f(x,y)+f(y,x)\big)$ and applying the APSF on summations over $n_1$, we obtain:
\begin{eqnarray}\label{Term1.massless.A1}
  T_1&=&\frac{-\lambda}{32a_1a_2}\bigg[\frac{1}{2}\sum_{n_1,n_2=1}^{\infty}\frac{1}{\sqrt{\frac{n_1^2\pi^2}{a^2_1}+\frac{n_2^2\pi^2}{a_2^2}}}
  +\{a_1\leftrightarrow a_2\}\bigg]^2\hspace{7cm}\nonumber \\
  &=& \frac{-\lambda}{32a_1a_2}\bigg[\frac{1}{2}\sum_{n_2=1}^{\infty}\bigg(\frac{-a_2}{2n_2\pi}
  +\int_{0}^{\infty}\frac{dx}{\sqrt{(\frac{x\pi}{a_1})^2+(\frac{n_2\pi}{a_2})^2}}+B_{1}(0;a_1,a_2)\bigg)+\{a_1\leftrightarrow a_2\}\bigg]^2,
\end{eqnarray}
where
\begin{eqnarray}\label{branch.cut.massless.}
 \small B_{1}(0;x,y)=i\int_{0}^{\infty}\Bigg(\frac{1}{\sqrt{(\frac{i\pi t}{x})^2+(\frac{n_2\pi}{y})^2}}-\frac{1}{\sqrt{(\frac{-i\pi t}{x})^2+(\frac{n_2\pi}{y})^2}}\Bigg)\frac{dt}{e^{2\pi t}-1}
  =\frac{2x}{\pi}\int_{1}^{\infty}\frac{d\eta}{\sqrt{\eta^2-1}(e^{2\pi n_2 x\eta/y}-1)},
\end{eqnarray}
is the Branch-cut term of APSF. Re-applying the APSF on all terms in the rhs of Eq.\,\eqref{Term1.massless.A1}, utilizing the BSS and CRT yield all infinite contribution in Eq.\,\eqref{Term1.massless.A1} would be removed and there is not remained any contributions from $T_1$ in the final Casimir energy expression.
\par
For the second and third terms of Eq.\,\eqref{All.Four.Terms} after applying the APSF we have:
\begin{eqnarray}\label{term2&3.massless}
 T_2+T_3=\frac{-\lambda}{32a_1a_2}\frac{1}{2}\sum_{n_1=1}^{\infty}\Bigg[\frac{-a_1}{2n_1\pi}
 +\underbrace{\int_{0}^{\infty}\frac{dx}{\sqrt{\frac{n_1^2\pi^2}{a_1^2}+\frac{x^2\pi^2}{a_2^2}}}}_{I_1(a_1,a_2)}+B_{1}(0;a_2,a_1)\Bigg]^2+\{a_1\leftrightarrow a_2\}\hspace{3.7cm}\nonumber\\
 =\frac{-\lambda}{32a_1a_2}\frac{1}{2}\sum_{n_1=1}^{\infty}\Bigg[\frac{a_1^2}{4\pi^2n_1^2}+I_1^2(a_1,a_2)-\frac{a_1}{\pi n_1}I_1(a_1,a_2)-\frac{a_1}{\pi n_1}B_{1}(0;a_2,a_1)\hspace{4.2cm}\nonumber\\
 +B^2_{1}(0;a_2,a_1)+2I_1(a_1,a_2)B_{1}(0;a_2,a_1)\Bigg]+\{a_1\leftrightarrow a_2\}.\hspace{4.1cm}
\end{eqnarray}
By re-applying the APSF on the first three terms in the rhs of Eq.\,\eqref{term2&3.massless}, utilizing the BSS and CRT, all contributions of these terms will be removed and there does not remain any contribution from them in the final expression. Therefore, we obtain:
\begin{eqnarray}
T_2+T_3=\frac{-\lambda}{32a_1a_2}\frac{1}{2}\sum_{n_1=1}^{\infty}\Bigg\{-\frac{a_1}{\pi n_1}B_{1}(0;a_2,a_1)+B^2_{1}(0;a_2,a_1)+2I_1(a_1,a_2)B_{1}(0;a_2,a_1)\Bigg\}+\{a_1\leftrightarrow a_2\},
\end{eqnarray}
where $I_1(x,y)$  is still divergent for every value of $n_1$. Using the BSS and CRT it can be shown that the remaining finite expression for this term becomes: $\frac{y}{\pi}\ln2$. Therefore the final expression for $T_2+T_3$ is obtained as:
\begin{eqnarray}
T_2+T_3=\frac{-\lambda}{32a_1a_2}\frac{1}{2}\sum_{n_1=1}^{\infty}\Bigg\{-\frac{a_1}{\pi n_1}B_{1}(0;a_2,a_1)+B^2_{1}(0;a_2,a_1)+\frac{2a_2\ln2}{\pi}B_{1}(0;a_2,a_1)\Bigg\}+\{a_1\leftrightarrow a_2\},
\end{eqnarray}
For the last term of Eq.\,\eqref{All.Four.Terms} in the limit $m=0$ after applying the APSF we obtain:
\begin{eqnarray}\label{Term4.1}
T_4=\frac{-\lambda}{32a_1a_2}\frac{1}{4}\sum_{n_1=1}^{\infty}\Bigg[\frac{a_1^2}{2\pi^2n_1^2}
+\underbrace{\int_{0}^{\infty}\frac{dx}{\frac{n_1^2\pi^2}{a_1^2}+\frac{x^2\pi^2}{a_2^2}}}_{=\frac{a_1a_2}{2\pi n_1}}\Bigg],
\end{eqnarray}
where the value of the Branch-cut term is zero. To convert the remaining summations given in Eq.\,\eqref{Term4.1} into the integral form, we reused the APSF on the summations of Eq.\,\eqref{Term4.1}. Next, using the BSS and CRT, all divergent expressions would be removed and there will not be remained any contribution for $T_4$. Finally, by integrating the aforementioned four parts and calculating the limits defined in Eq.\,\eqref{BSS.CAS.Def.}, the radiative correction to the Casimir energy of massless scalar field in the rectangle is obtained. This final expression was reported in Eq.\,\eqref{Massless.Casimir.Expression.}.

\acknowledgments
The Author would like to thank the research office of Semnan Branch, Islamic Azad University for the financial support.


\begin{thebibliography}{99}

\bibitem{h.b.g.}
     H. B. G. Casimir, Proc. Kon. Nederl. Akad. Wet. {\bf 51} (1948) 793.

\bibitem{Sparnaay.M.J.}
     M. J. Sparnaay,\emph{Measurements of attractive forces between flat plates}, Physica {\bf 24} (1958) 751.

\bibitem{quantum.field.theory.1}
     K. A. Milton, \emph{The Casimir Effect: Physical Manifestations of Zero-Point Energy}, (World
     Scientific Publishing Co. 2001);\\
     L. Shahkarami, A. Mohammadi and S. S. Gousheh, \emph{Casimir energy for a coupled fermion-soliton system}, JHEP \textbf{11} (2011) 140;\\
     N. R. Khusnutdinov and R. N. Kashapov, \emph{Casimir Effect For a Colection of Parallel Conducting Surfaces}, Theo. Math. Phys.  {\bf183} (2015) 491;\\
     J. Lorenzen and D. Martelli, \emph{Comments on the Casimir energy in supersymmetric field theories}, JHEP \textbf{07}(2015) 001.
\bibitem{condensed.matter.1}
    A. Edery, Phys. Rev. D \textbf{75} (2007) 105012.

\bibitem{atomic.molecular.1}
    M. A. Braun, \emph{Casimir Energy of The Quantum Field in a Dispersive and Absorptive Medium}, Theo. Math. Phys. {\bf175} (2013) 771.

\bibitem{astro.physics.1}
    G. Mahajan, S. Sarkar and T. Padmanabhan, Phys. Lett. B \textbf{641} (2006) 6.

\bibitem{Mathematical.physics.1}
     A. Romeo, K.A. Milton, \emph{Casimir energy for a purely dielectric cylinder by the mode summation method}, Phys. Lett. B \textbf{621} (2005) 309.

\bibitem{Bordag.et.al}
     M. Bordag, D. Robaschik, E. Wieczorek, Ann. Phys. (N.Y.) {\bf 165} (1985) 192;\\
     M. Bordag and J. Lindig, \emph{Radiative correction to the Casimir force on a sphere},Phys. Rev. D {\bf58} (1998) 045003;\\
     M. Bordag, K. Scharnhorst, Phys. Rev. Lett. {\bf 81} (1998) 3815.

\bibitem{RC.free.counterterms.}
     L.C. de Albuquerque, Phys. Rev. D 55 (1997) 7754;\\
     M. Bordag, U. Mohideen, V.M. Mostepanenko, Phys. Rep. 353 (2001) 1.
\bibitem{13-20reza}
    F.A. Baron, R.M. Cavalcanti, C. Farina, Nucl. Phys. B (Proc. Suppl.) 127 (2004) 118.
\bibitem{cavalcanti.}
    F.A. Baron, R.M. Cavalcanti, C. Farina, hep-th/0312169.
\bibitem{21-reza}
    N. Graham, R. Jaffe, Weigel, Int. J. Mod. Phys. A 17 (2002) 864.
\bibitem{22-reza}
    N. Graham, R.L. Jaffe, V. Khemani, M. Quandt, O. Schroeder, H. Weigel, Nucl. Phys. B 677 (2004) 379.

\bibitem{SS.Gousheh.etal}
    R. Moazzemi, M. Namdar and S. S. Gousheh, JHEP {\bf 09}(2007)029;\\
    R. Moazzemi, A. Mohammadi, S. S. Gousheh, Eur. Phys. J. C {\bf 56} (2008) 585;\\
    R. Moazzemi, S. S. Gousheh, Phys. Lett. B {\bf 658} (2008) 255.

\bibitem{mode.summation.}
     A. Romeo, K. A. Milton, Phys.Lett. B \textbf{621} (2005) 309;\\
     K. A. Milton, A. V. Nesterenko, V. V. Nesterenko, \emph{Mode-by-mode summation for the zero point electromagnetic energy of an infinite cylinder}, Phys.Rev.D \textbf{59} (1999) 105009;\\
     I. H. Brevik, V. V. Nesterenko, I. G. Pirozhenko, \emph{Direct mode summation for the Casimir energy of a solid ball}, J. Phys. A \textbf{31} (1998) 8661.

\bibitem{Green.function.}
     K. A. Milton, L. L. Deraad, and J. Schwinger, \emph{Casimir self-stress on a perfectly conducting spherical shell}, Ann. Phys. (N.Y.) \textbf{115} (1978) 388;\\
    J. Baacke and Y. Igarashi, \emph{Casimir energy of confined massive quarks}, Phys. Rev. D \textbf{27} (1983) 460.

\bibitem{schwinger.source.theory.}
    J. Schwinger, L. L. de Raad Jr., and K. A. Milton, \emph{Casimir effect in dielectrics}, Ann. Phys. \textbf{115} (1978) 1;\\
    J. Schwinger, \emph{Casimir Effect in source theory}, Lett. Math. Phys. \textbf{1} (1975) 43;\\
    K. A. Milton, L. L. de Raad Jr., and J. Schwinger, \emph{Casimir self-stress on a perfectly conducting spherical shell}, Ann. Phys., \textbf{115} (1978) 388.
\bibitem{sphere.man}
     M. A. Valuyan and S. S. Gousheh, Int. J. Mod. Phys. A \textbf{25} (2010) 1165.
     R. Saghian, M. A. Valuan, A. Seyedzahedi and S. S. Gousheh, Int. J. Mod. Phys. A \textbf{27} (2012) 1250038.
\bibitem{boyer.}
     T. H. Boyer, Phys. Rev. \textbf{174} (1968) 1764.

\bibitem{2dim.valuyan}
    S. S. Gousheh, R. Moazzemi and  M. A. Valuyan, Phys. Lett. B \textbf{681} (2009) 477.

\bibitem{curved.valuyan}
    M. A. Valuyan, \emph{Radiative correction to the Casimir energy for massive scalar field on a spherical surface}, Mod. Phys. Lett. A \textbf{32} (2017) 1750128;\\
    M. A. Valuyan, \emph{Casimir Energy Calculation for Massive Scalar Field on Spherical Surface: An Alternative Approach}, Canadian J. Phys. \textbf{96} (2018) 722;\\
    F. Br$\ddot{u}$nner, D. Regaladob, and V. P. Spiridonovd, JHEP \textbf{07}(2017)041;\\
    L.P. Teo, \emph{Casimir interaction between spheres in $D+1$ dimensional Minkowski spacetime}, JHEP \textbf{05}(2014)016;\\
    B. Assel, D. Cassani, L. D. Pietro, Z. Komargodski, J. Lorenzena and D. Martellia, \emph{The Casimir energy in curved space and its supersymmetric counterpart}, JHEP \textbf{07}(2015)043.

\bibitem{waveguide.man}
      M. A. Valuyan, R. Moazzemi, and S. S. Gousheh, \emph{A direct approach to the electromagnetic Casimir energy in a rectangular waveguide}, J. Phys. B: At. Mol. Opt. Phys. \textbf{41} (2008) 145502.

\bibitem{wolfram.locuz}
     W. Lukosz, \emph{Electromagnetic zero-point energy and radiation pressure for a rectangular cavity}, Physica \textbf{56} (1971) 109;\\
     J. Ambj{\o}rn and S. Wolfram, 1983 \emph{Properties of the vacuum: I. Mechanical and thermodynamic} Ann. Phys., NY \textbf{147}  (1983) 1.

\bibitem{Generalized.Abel.Plana.Saharian}
     A.A. Saharian, \emph{The Genetalized Abel-Plana Formula: Applications To Bessel Functions And Casimir Effect} IC/2007/082 (2000)
     [{ \tt hep-th/0002239 v1}].

\end{thebibliography}
\end{document}